%% file: main.tex
\documentclass[sigconf,screen]{acmart}
\usepackage[ruled,linesnumbered]{algorithm2e}

\input{tex/preamble}

\copyrightyear{2023} 
\acmYear{2023} 
\setcopyright{acmlicensed}\acmConference[CF '23]{20th ACM International Conference on Computing Frontiers}{May 9--11, 2023}{Bologna, Italy}
\acmBooktitle{20th ACM International Conference on Computing Frontiers (CF '23), May 9--11, 2023, Bologna, Italy}
\acmPrice{15.00}
\acmDOI{10.1145/3587135.3592199}
\acmISBN{979-8-4007-0140-5/23/05}

\begin{document}
\settopmatter{printfolios=true}
\input{tex/title}
\input{tex/abstract}



\maketitle

\input{tex/intro}

\input{tex/motivation}

\input{tex/design}

\input{tex/implementation}

\input{tex/methodology}

\input{tex/eval}

\input{tex/related}

\input{tex/conclusion}

\bibliographystyle{ACM-Reference-Format}
\bibliography{ref}

\end{document}

%% file: tex/preamble.tex
\usepackage[normalem]{ulem}

\usepackage{color}
\usepackage{threeparttable}
\usepackage{float}
\usepackage{amsmath,amsfonts}
\usepackage{graphicx}
\usepackage{textcomp}
\usepackage{multirow}
\usepackage{colortbl}
\usepackage[ruled]{algorithm2e}
\usepackage{diagbox}
\usepackage{mathrsfs}
\usepackage{booktabs}
\usepackage{hhline}
\usepackage{array,ragged2e}
\usepackage{tabularx}
\usepackage{algorithmic}
\usepackage{xcolor}
\usepackage{lipsum}

\usepackage{verbatim}
\usepackage{enumitem}
\usepackage[colorinlistoftodos,prependcaption,textsize=small]{todonotes}
\usepackage[ruled]{algorithm2e}
\usepackage{multicol,multicap}

\usepackage{subfig}
\usepackage{listings}
\usepackage{fancyhdr} 

\usepackage{soul}

\usepackage{tikz}

\definecolor{bg}{rgb}{0.96,0.96,0.96}

\newcommand{\Fig}[1]{Fig.~\ref{#1}}

\newcommand{\Tbl}[1]{Tbl.~\ref{#1}}
\newcommand{\Sec}[1]{Sec.~\ref{#1}}

\newcommand{\Algo}[1]{Algo.~\ref{#1}}

\renewcommand{\paragraph}[1]{\vspace*{.1cm}\noindent\textbf{#1}\hspace*{.1cm}}

\newcommand{\proj}[1]{$AdaptGear$}

\def\BibTeX{{\rm B\kern-.05em{\sc i\kern-.025em b}\kern-.08em
    T\kern-.1667em\lower.7ex\hbox{E}\kern-.125emX}}



%% file: tex/title.tex
\title{AdaptGear: Accelerating GNN Training via Adaptive Subgraph-Level Kernels on GPUs}

\author{Yangjie Zhou}
\affiliation{%
  \institution{Shanghai Jiao Tong University}
  \city{Shanghai}
  \country{China}
}
\email{yj_zhou@sjtu.edu.cn}

\author{Yaoxu Song}\affiliation{
    \institution{Shanghai Jiao Tong University}
    \city{Shanghai}
    \country{China}
}\email{Richard_K@sjtu.edu.cn}

\author{Jingwen Leng}
\affiliation{
    \institution{Shanghai Jiao Tong University}
    \institution{Shanghai Qi Zhi Institusion}
    \city{Shanghai}
    \country{China}
}\email{leng-jw@cs.sjtu.edu.cn}
\authornotemark[1]

\author{Zihan Liu
}\affiliation{
    \institution{Shanghai Jiao Tong University}
    \city{Shanghai}
    \country{China}
}\email{altair.liu@sjtu.edu.cn}

\author{Weihao Cui}
\affiliation{
    \institution{Shanghai Jiao Tong University}
    \city{Shanghai}
    \country{China}
}\email{weihao@sjtu.edu.cn}

\author{Zhendong Zhang}
\affiliation{
    \institution{Shanghai Qi Zhi Institute}
    \city{Shanghai}
    \country{China}
}\email{zhangzd@sqz.ac.cn}

\author{Cong Guo}
\affiliation{%
    \institution{Shanghai Jiao Tong University}
    \institution{Shanghai Qi Zhi Institusion}
    \city{Shanghai}
    \country{China}
}\email{guocong@sjtu.edu.cn}

\author{Quan Chen}
\affiliation{%
    \institution{Shanghai Jiao Tong University}
    \city{Shanghai}
    \country{China}
}\email{chen-quan@cs.sjtu.edu.cn}

\author{Li Li}
\affiliation{%
    \institution{Shanghai Jiao Tong University}
    \city{Shanghai}
    \country{China}
}\email{lilijp@sjtu.edu.cn}

\author{Minyi Guo}
\affiliation{
    \institution{Shanghai Jiao Tong University}
    \city{Shanghai}
    \country{China}
}\email{guo-my@cs.sjtu.edu.cn}
\authornote{Jingwen Leng and Minyi Guo are the corresponding authors.}

%% file: tex/abstract.tex
\begin{abstract}

Graph neural networks (GNNs) are powerful tools for exploring and learning from graph structures and features. 
As such, achieving high-performance execution for GNNs becomes crucially important.
Prior works have proposed to explore the sparsity (i.e., low density) in the input graph to accelerate GNNs, which uses the full-graph-level or block-level sparsity format.
We show that they fail to balance the sparsity benefit and kernel execution efficiency.
In this paper, we propose a novel system, referred to as \proj{}, that addresses the challenge of optimizing GNNs performance by leveraging kernels tailored to the density characteristics at the subgraph level.
Meanwhile, we also propose a method that dynamically chooses the optimal set of kernels for a given input graph. Our evaluation shows that \proj{} can achieve a significant performance improvement, up to $6.49 \times$ ($1.87 \times$ on average), over the state-of-the-art works on two mainstream NVIDIA GPUs across various datasets.

\end{abstract}

%% file: tex/intro.tex
\section{Introduction}
\label{sec:intro}

Optimizing graph neural networks (GNNs) performance is a vital task of great interest to both academia and industry. 
As GNNs have demonstrated success in extending deep learning to graph structures and features, research in this area has advanced rapidly in recent years. Various variants of GNN models have been designed and explored, leading to significant breakthroughs in fields such as chemistry~\cite{MPNN}, neurology~\cite{neuroscience}, anomaly detection~\cite{Spam, CARE-GNN}, and social networks or recommendations~\cite{GraphRec, UltraGCN}. 
As graph-structured data continues to grow, it is becoming increasingly imperative to optimize GNN performance to enable real-time analysis and decision-making~\cite{gnn_survey, wu2020graph}.

One of the important factors in optimizing the performance of GNNs is to understand and utilize the density/sparsity nature of their input graphs~\cite{GOP-Sp,gnn_compute_survey}. 
Real-world graphs commonly exhibit community-based structures~\cite{fortunato2010community,lancichinetti2008benchmark,newman2013spectral}, which can be identified using existing community-based ordering tools by grouping similar vertices together in ordinal order~\cite{METIS,rabbit,graph_reorder,LLP}. This can result in variability in density distribution within the adjacency matrix of a single graph, with higher density on the diagonal reflecting the edge connectivity within a community, and lower density in other locations reflecting the edge connections between communities. 
Additionally, different input graph datasets can have distinct density characteristics, with the difference in density between different graphs potentially reaching several orders of magnitude~\cite{ogb}.
Hence, it is imperative to consider these density characteristics while optimizing the performance of GNNs to ensure efficiency in training and deployment.

Previous studies on exploiting the graph sparsity (i.e., low density) on the modern parallel GPU platform can be divided into two categories based on the granularity of kernel mapping. 
The first category, referred to as full-graph-level kernel mapping, employs a single optimized kernel for the entire graph~\cite{GNNAdvisor,Ge-SPMM,NeuGraph}. 
The second category is referred to as block-level kernel mapping. The input graph's adjacency matrix is divided into blocks during preprocessing. The optimal execution mode for each block is determined based on its density. After computation for each block is finished, the results are combined for blocks that correspond to the same set of vertices~\cite{PCGCN}.
However, these approaches are insufficient to fully utilize the graph sparsity for accelerating GNNs. 
Specifically, the full-graph-level kernel mapping disregards the intra-graph density distribution.
The block-level kernel mapping incurs additional overhead in runtime due to the kernel launch and result combination processes.

In this paper, we introduce \proj{}, a novel system that addresses the challenge of optimizing the performance of GNNs by leveraging the unique characteristics of graph density distribution with minimal runtime overhead. 
The system starts by decomposing the input graph into subgraphs corresponding to intra-community and inter-community based on graph community features in a preprocessing stage.
We then employ two key components for computational optimization.
The first component, subgraph-level customized kernels, offers diverse kernel formats tailored to the specific characteristics with varying densities of individual subgraphs, resulting in more efficient utilization of computational resources and a more efficient training process.
The second component, the adaptive selector, selects the appropriate kernel based on a feedback-driven approach during runtime. This ensures that the optimal kernel is used for each specific input graph, further improving computational performance and avoiding runtime overhead. 
Based on the innovative design, \proj{} offers a novel and effective approach to optimize the performance of GNNs by utilizing the unique characteristics of graph density distribution.

Overall, our work makes the following contributions:
\begin{itemize}[ itemsep=0pt, parsep=0pt, labelsep=5pt, 
    leftmargin=*, topsep=0pt,partopsep=0pt]
    \item We conduct a detailed analysis of the density distribution of both intra- and inter-graphs in order to understand their impact on the sparsity-based GNN execution approaches.
    \item We propose a set of subgraph-level customized kernels, which are tailored to the density characteristics of specific subgraphs to improve GNNs' computational efficiency.
    \item We introduce an adaptive selector that can determine the best-performing kernel based on a feedback-driven approach. This ensures that the optimal kernel is used for each specific graph, further improving performance and reducing runtime overhead.
    \item We propose \proj{} to integrate the above optimizations and evaluate it against state-of-the-art techniques to demonstrate its effectiveness in optimizing GNN performance. Our results show that \proj{} significantly improves performance, with an average improvement of $1.87 \times$ compared to existing methods.
\end{itemize}

%% file: tex/motivation.tex
\section{Background and Motivation}
\label{sec:motivation}

This section provides a brief background of graph neural networks (GNNs) and their distinction from traditional deep neural networks (DNNs). 
We show that GNNs incorporate graph-related inputs so that their execution efficiency varies depending on the storage format of the input graph.
We then examine the density distribution of both intra- and inter-graphs, which highlights the need for subgraph-level adaptive kernels in optimizing GNN performance.

\subsection{Graph Neural Networks}
\label{sec:background:gnn}

Graph neural networks (GNNs) have recently gained significant attention in both academia and industry for their ability to effectively learn and infer on graph-structured data in non-Euclidean spaces~\cite{graph_ml_survey,gnn_survey}. 
As such, achieving high-performance execution of GNNs has become an important research topic~\cite{Gnnmark,gnn_survey,gnn_compute_survey}. 
The input of a GNN model is a $d$-dimensional vector representation for each vertex in the input graph, known as an embedding. These embeddings are designed to have similar values for vertices with similar properties, such as similar subgraph structures, to facilitate efficient reasoning about graph-related problems~\cite{GCN,Sage}. 
To generate these embeddings, GNNs combine feature transformation methods from DNNs with graph-based operations in graph processing that propagate and aggregate information throughout the graph structure.

\begin{figure}[t]
    \centering
    \includegraphics[width=0.98\linewidth]{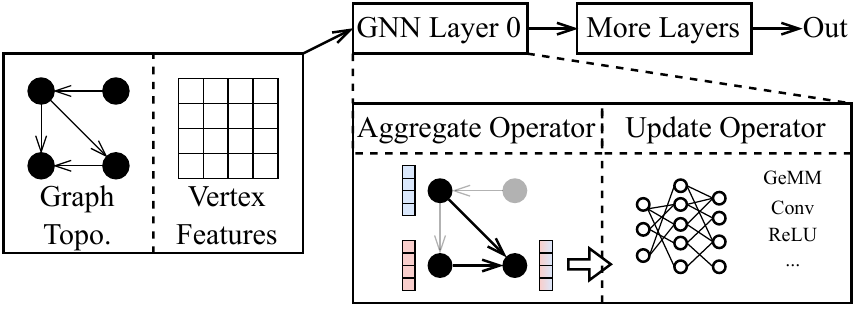}
    \caption{GNN general computation flow.}
    \label{fig:GNN_Flow}
    \vspace{-0.2cm}

\end{figure}

\paragraph{GNN Computation.} 
The computation flow of GNNs is illustrated in \Fig{fig:GNN_Flow}.
The input of a GNN model comprises both the topological structure of the input graph and the dense feature embeddings associated with each vertex. The computation within each GNN layer is composed of two primary types of operations, as represented by the following equations:
$$
\begin{aligned}
    a_{v}^{(k)} &=\operatorname{Aggregate}^{(k)}\left(\left\{h_{u}^{(k-1)} \mid u \in \mathcal{N}(v)\right\}\right) \\
    h_{v}^{(k)} &=\operatorname{Update}^{(k)}\left(h_{v}^{(k-1)}, a_{v}^{(k)}\right)
\end{aligned}
$$
where $h_{v}^{k}$ represents the feature vector of vertex $v$ at the $k$-th layer. The \texttt{Aggregate} function performs the aggregation of multiple feature vectors from adjacent vertices into a single feature vector using various specific operators, such as \texttt{max}, \texttt{mean}, and \texttt{sum}, which are referred to as \texttt{aggregate-max}, \texttt{aggregate-mean}, and \texttt{aggregate-sum}, respectively. The \texttt{Update} function utilizes neural network operations, such as a multilayer perceptron (MLP), to transform each vertex's feature vector into a new feature vector.

\begin{figure}[t]
    \centering
    \subfloat[]{
        \includegraphics[width=0.6\linewidth]{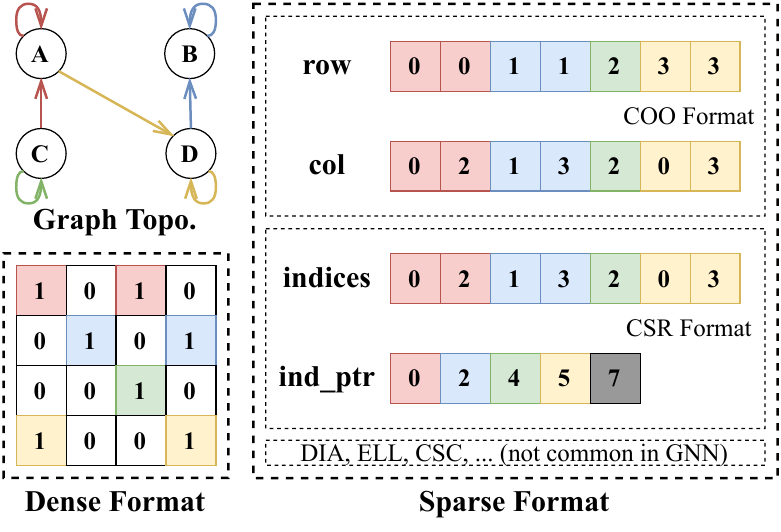}
        \label{fig:graph_format}
    }
    \subfloat[]{
        \includegraphics[width=0.38\linewidth]{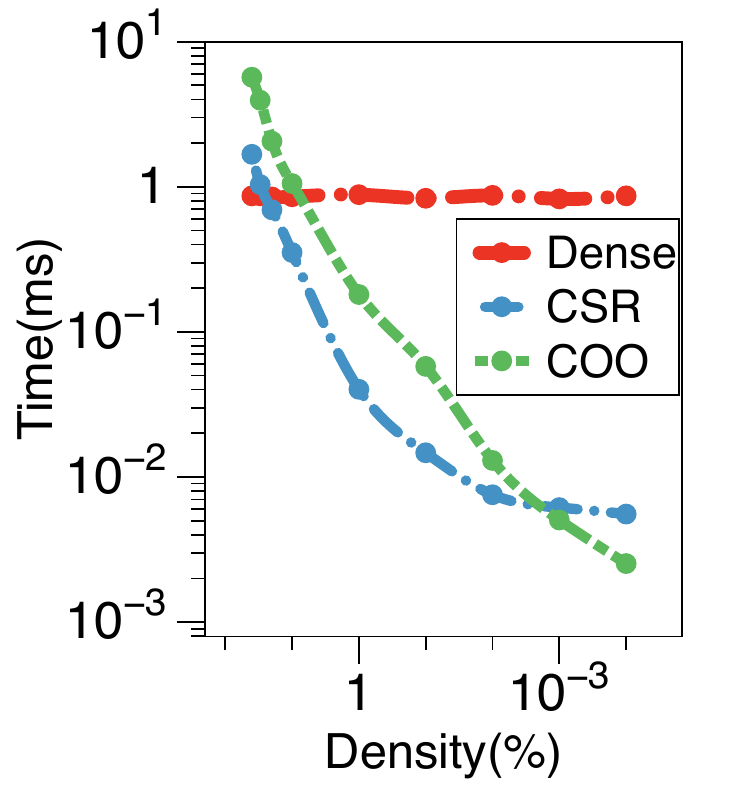}
        \label{fig:pubmed_sparse}
    }

    \caption{\small  (a) Graph format example: Dense/CSR/COO. (b) Performance comparison of different format for the \texttt{aggregate-sum} operator in GCN first layer with Pubmed dataset on A100 GPU.}

\end{figure}

\paragraph{Graph Format and Execution Scheduling.}
As a distinct feature that sets GNNs apart from traditional DNNs, GNNs' input data include graph topology information.
This topological data can be represented in various formats, such as the dense format, which utilizes a 2-dimensional array to represent the adjacency matrix of the graph, with a value of 1 indicating a connection between the corresponding source and destination vertices and 0 indicating otherwise. The dense format is convenient for continuous memory accesses but space-inefficient due to the presence of the large number of 0s in the graph.
To reduce the storage overhead, GNN systems commonly employ sparse graph formats, such as compressed sparse row (CSR) and coordinate format (COO). 
The CSR format contains two arrays: the row pointer array and the index columns array.
The row pointer array stores the indices of the start of each row in the columns array, while the columns array stores the column indices of the non-zero elements in the matrix. 
These two arrays can be used together to reconstruct the graph topology.
The COO format stores an array of tuple for non-zero elements. 
Each tuple represents an edge of the graph, with the row and column indices representing the destination and source vertices of the edge, respectively.
\Fig{fig:graph_format} shows examples of these different dense and sparse formats.

The choice of graph storage format not only affects the physical organization of the data, but also significantly impacts the scheduling approach that parallelizes the GNNs' execution on GPUs. Specifically, the dense format treats a graph-related operator as a dense operator, while the CSR and COO formats correspond to vertex-parallel~\cite{Ge-SPMM} and edge-parallel~\cite{gnn_understanding} approaches, respectively. Vertex-parallelism is achieved by assigning each thread to a specific vertex and processing all of its associated edges sequentially. 
In comparison, edge-parallelism is achieved by assigning each thread to a specific edge and processing these computations in parallel.

We show that the optimal graph format choice depends on the specific characteristics of the input graph and requires careful consideration.
In order to experimentally analyze how this choice is made, we generate input graphs with various densities using RMAT~\cite{chakrabarti2004r} tool. 
We adjust the number of edges with a fixed vertex size of 19717, which is the size of Pubmed dataset~\cite{Gnnmark}. 
Using an A100 GPU~\cite{a100_whitepaper}, we compare performance results for these graphs with different sparsity using CSR, COO, and dense data formats, performing \texttt{aggregate-sum} operations. The results shown in \Fig{fig:pubmed_sparse} indicate that the dense format has optimal execution efficiency at high density, CSR performs optimally as density decreases, and COO becomes the optimal solution at low density.

\subsection{Intra-Graph Density Analysis}
\label{subsec:motivation:intra-graph}

\begin{figure}[t]
    \centering
    \subfloat[]{
        \includegraphics[width=0.5\linewidth]{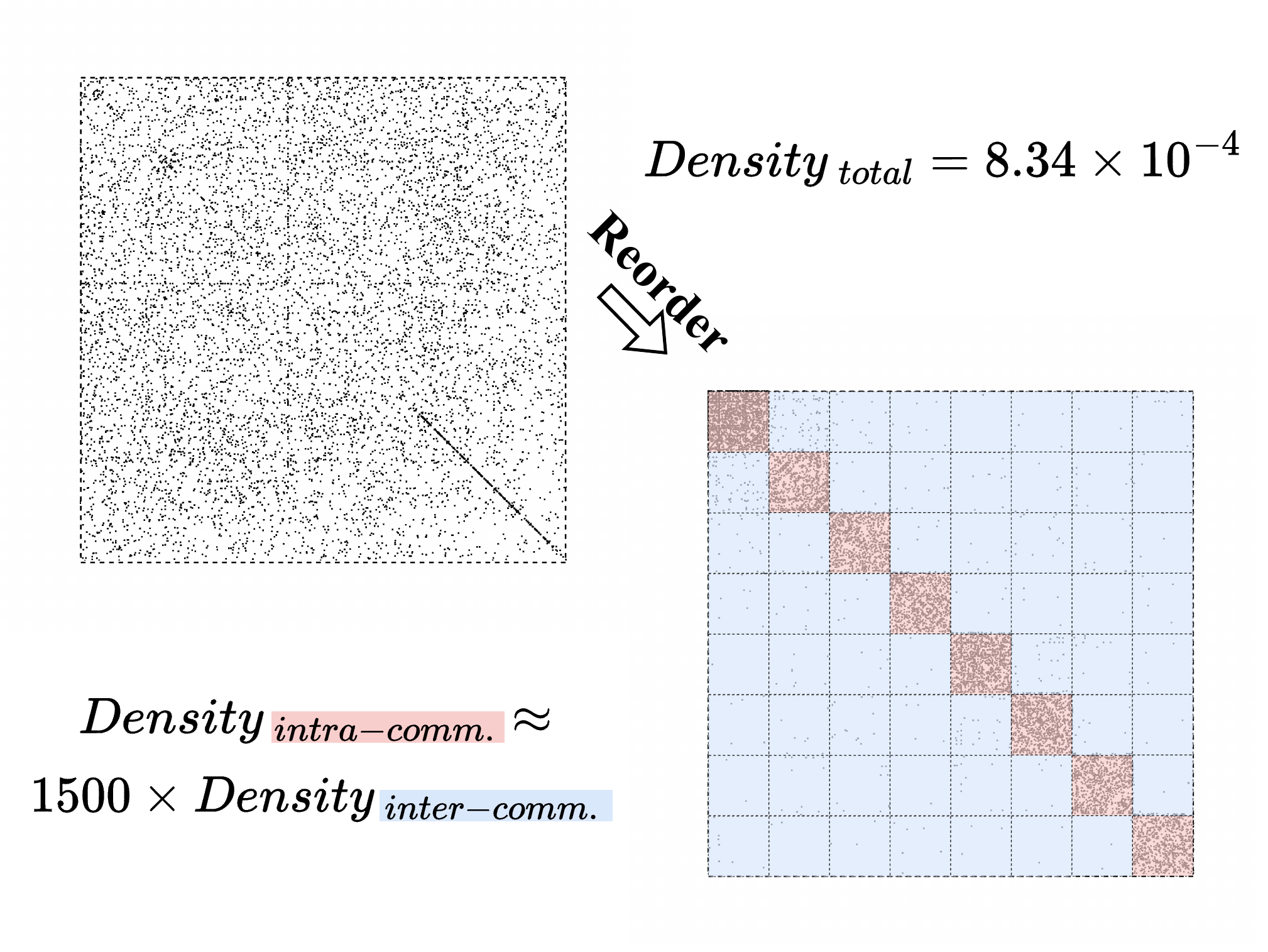}
        \label{fig:reorder_density}
    }
    \subfloat[]{
        \includegraphics[width=0.5\linewidth]{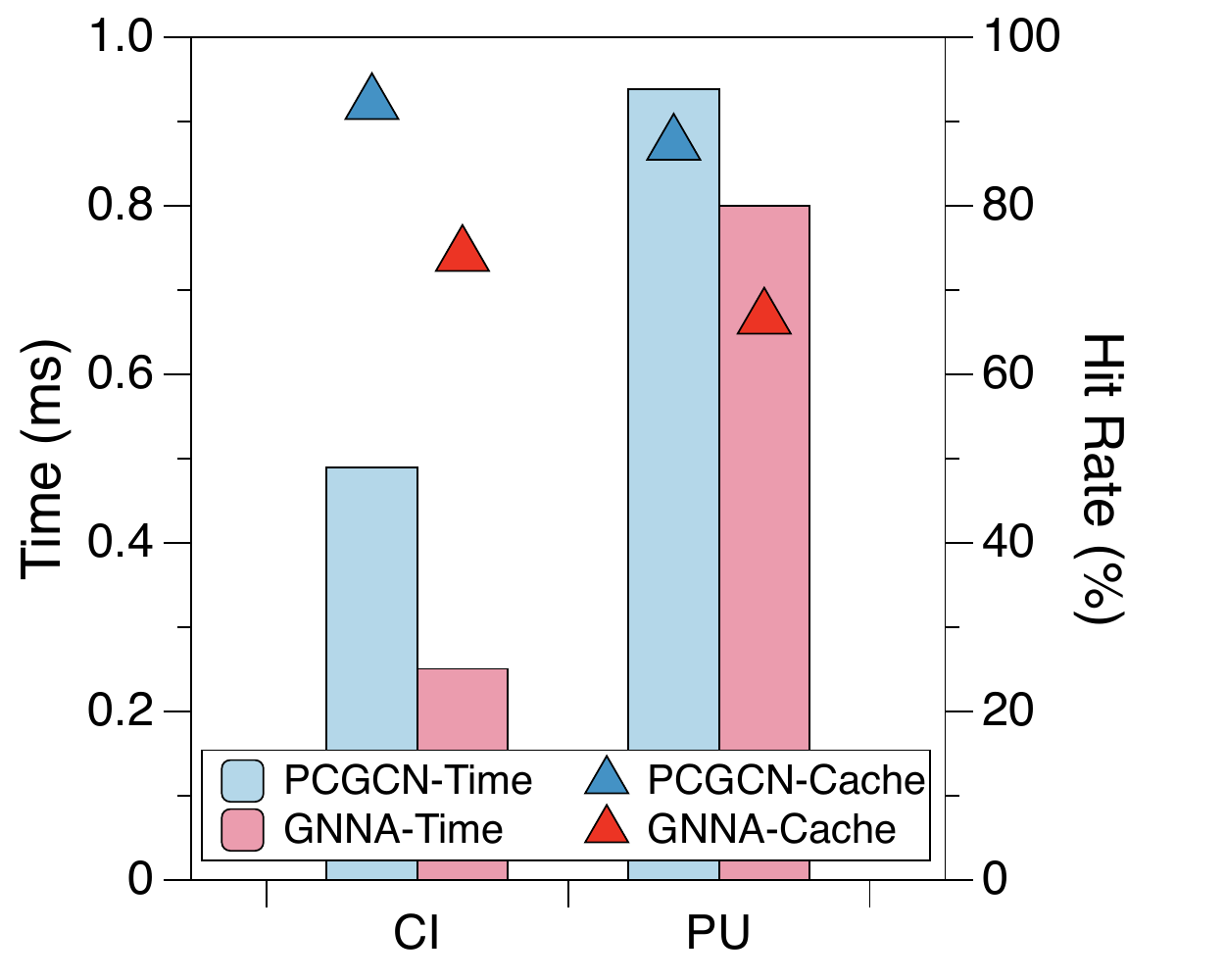}
        \label{fig:pcgcn_gnna_kernel_mapping}
    }
    
    \caption{\small  (a) The impact of commmunity-based reordering on the adjacency matrix of  Citeseer dataset. (b) Comparison of performance and L2 cache hit rate for the \texttt{aggregate-sum} operator in GCN first layer with Citeseer and Pubmed datasets on A100 GPU.}

\end{figure}

Real-world graphs commonly exhibit community-based structures~\cite{fortunato2010community,lancichinetti2008benchmark,newman2013spectral}, which can be identified using existing community-based ordering tools~\cite{METIS, rabbit} by grouping similar vertices together in ordinal order.
Each graph community has a group of vertices with strong intra-community connections, but weak connections to other vertices. However, the ordinal numbers of an original graph are often randomly assigned. 
Community-based reordering aims to reorder the vertices of a graph such that neighboring vertices belong to the same community.
This reordering also leads to a differentiated density distribution for different subgraphs within a graph.
This phenomenon is illustrated shown in \Fig{fig:reorder_density}, through an examination of the Citeseer dataset~\cite{giles1998citeseer} using METIS~\cite{METIS}, a commonly utilized community-based reordering tool.
The distribution of the adjacency matrix before reordering is observed to be random and irregular~\cite{graph_reorder}. However, after reordering, the distribution of the adjacency matrices corresponding to intra- and inter-community edges exhibits distinct characteristics.
These edges correspond to the intra- and inter-community subgraphs, respectively.
These two subgraphs display a significant difference in density, with the density of intra-communities edges (i.e., on the diagonal) higher than the density of inter-communities edges.

Previous research has explored using the graph topology obtained after reordering for high-performance GNN computation. This research can be classified into two categories based on the granularity of kernel mapping.
The first category, referred to as full-graph-level kernel mapping, implements a static optimization kernel for the entire graph~\cite{GNNAdvisor,NeuGraph}. The second category, referred to as block-level kernel mapping, invokes the computational kernel independently for each block of the adjacency matrix~\cite{PCGCN}. 
The full-graph-level mapping ignores the density distribution pattern brought by the community-based reordering of the adjacency matrix and treats it as a form of the orthogonal preprocessing optimization method. 
In contrast, the block-level kernel mapping selects the appropriate execution mode for each block based on its density at a fine-grained level, and it then merges the results of the blocks corresponding to the same set of vertices.

As typical examples of these two categories, we use GNNAdvisor~\cite{GNNAdvisor} and PCGCN~\cite{PCGCN} for performance analysis. We collected the execution time of the first layer of GCN and the L2 cache hit rate via nsight system profiler~\cite{nvidia_profiler} on A100 GPU. 
The results in \Fig{fig:pcgcn_gnna_kernel_mapping} reveal that while PCGCN achieves a higher cache hit rate, it incurs a longer execution time. This is due to the fact that PCGCN employs an overly fine-grained granularity of kernel mapping, which incurs additional runtime overhead of kernel launching and results merging.
Therefore, it is essential to identify an appropriate mapping granularity and leverage the distribution characteristics of the reordered adjacency matrix to enhance performance.

\begin{figure}[t]
    \centering
    \includegraphics[width=0.98\linewidth]{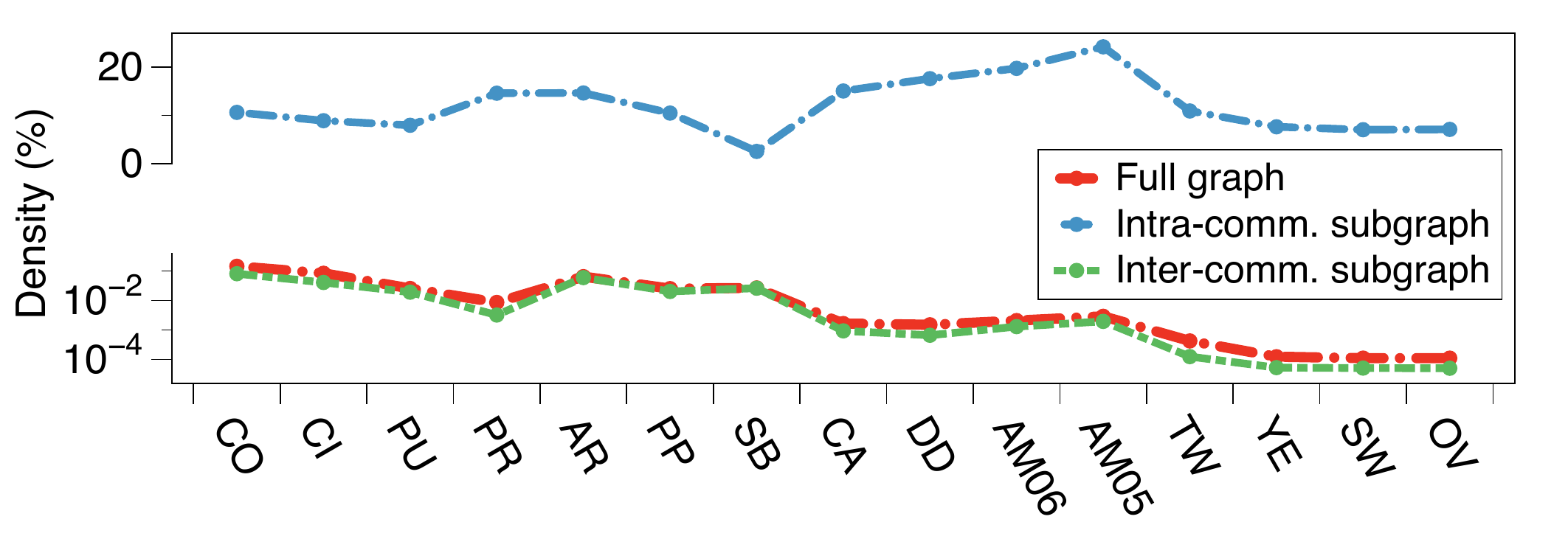}

    \caption{The average density of full, intra-community and inter-community subgraphs for different datasets.}
    \label{fig:alldata_distribution}
    
\end{figure}

\begin{table}[b] 
    \caption{Details of graph datasets used for evaluation.}
    \label{tbl:dataset}
    \center
    \resizebox{0.48\textwidth}{!}{
        \begin{tabular}{@{}lllll@{}}
            \toprule
        dataset                        & \#Vertex & \#Edge  & \#Feat & \#Class \\ \midrule
        cora (CO)                       & 2708     & 10556   & 1433   & 7       \\
        citeseer (CI)                   & 3327     & 9228    & 3703   & 6       \\
        pubmed (PU)                     & 19717    & 99203   & 500    & 3       \\
        PROTEINS\_full (PR)             & 43466    & 162088  & 29     & 2       \\
        artist (AR)                     & 50515    & 1638396 & 100    & 12      \\
        ppi (PP)                        & 56944    & 818716  & 50     & 121     \\
        soc-BlogCatalog (SB)            & 88784    & 2093195 & 128    & 39      \\
        com-amazon (CO)                 & 334863   & 1851744 & 96     & 22      \\
        DD                             & 334925   & 1686092 & 89     & 2       \\
        amazon0601 (AM06)               & 403394   & 3387388 & 96     & 22      \\
        amazon0505 (AM05)               & 410236   & 4878874 & 96     & 22      \\
        TWITTER-Real-Graph-Partial (TW) & 580768   & 1435116 & 1323   & 2       \\
        Yeast (YE)                      & 1710902  & 3636546 & 74     & 2       \\
        SW-620H (SW)                    & 1888584  & 3944206 & 66     & 2       \\
        OVCAR-8H (OV)                   & 1889542  & 3946402 & 66     & 2       \\ \bottomrule
        \end{tabular}
    }
\end{table}

\subsection{Inter-Graph Density Analysis}

In real-world graphs, density distribution variations manifest both at the intra-graph and inter-graph levels. 
In other words, different graphs have significant differences in density properties. 

To investigate these variations, we analyze 15 commonly used graph datasets as shown in \Tbl{tbl:dataset}.
The number of vertices and edges in each dataset is recorded to provide insight into the scale of the graph.
Additionally, the features and class sizes of the datasets vary, which could impact the computational complexity and memory usage of certain graph operations.

We quantify and analyze the variability of density distribution for different datasets after community-based reordering using the METIS algorithm. The size of each community is set to 16. 
As depicted in \Fig{fig:alldata_distribution}, the results confirm our previous analysis on intra-graph in \Sec{subsec:motivation:intra-graph} that there are distinctions in the density distributions of intra-community subgraphs and inter-community subgraphs.
Furthermore, the results demonstrate that these distinctions in characteristics also exist between different datasets.
As such, it can be inferred that using a fixed format for all datasets does not result in optimal performance improvement.

%% file: tex/design.tex
\section{Design of \proj{}}
\label{sec:design}

In this section, we present our training system, \proj{}, which efficiently exploits both the intra- and inter-graph sparsity to accelerate GNN training.
We first present an overview of our system and then present details for its two key components, which are the subgraph-level customized kernels and adaptive selector.

\subsection{Overview}

\begin{figure}[t]
    \centering
    \includegraphics[width=0.98\linewidth]{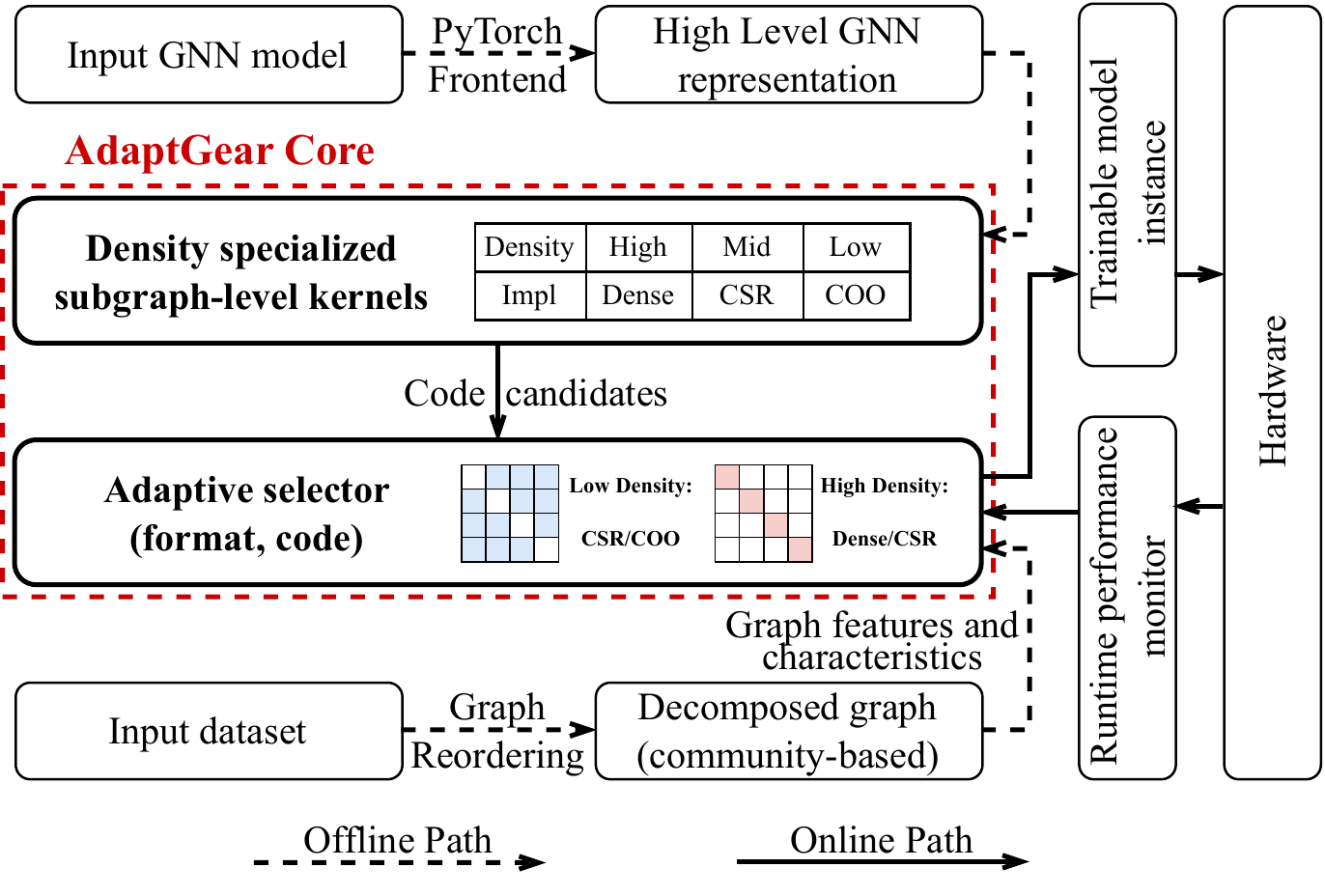}
    \caption{The design of \proj{}.}
    \label{fig:overview}
\end{figure}

\Fig{fig:overview} illustrates the overview of \proj{}, a GNN acceleration system that utilizes subgraph-level adaptive kernels. The input GNN model is represented using a PyTorch front-end, while the input graph dataset is decomposed under community-based reordering guidelines. The core design of \proj{} includes two modules: customized CUDA kernel templates and an adaptive code selector.

The first core component of \proj{} is the customization of CUDA kernels, which are designed to handle intra- and inter-graphs with varying density. 
Instead of using static kernels, \proj{} designs customized kernels for both intra-community and inter-community subgraphs with varying densities. 
This approach establishes a comprehensive strategy space, offering the potential for high-performance optimizations of GNNs.

The second core component of \proj{} is the adaptive code selector, which selects the optimal CUDA kernel template during runtime. 
By monitoring and profiling the performance of each subgraph kernel during the initial few iterations, the adaptive code selector selects the optimal CUDA kernel template that is best suited to the corresponding inputs.

\begin{table}[t]
    \caption{Comparison of existing graph operator acceleration methods with our proposed \proj{}.}
    \label{tbl:kernel_mapping}
    \resizebox{0.48\textwidth}{!}{
    \begin{tabular}{@{}cccc@{}}
    \toprule
    \begin{tabular}[c]{@{}c@{}}Kernel Mapping\\ Granularity\end{tabular} & \begin{tabular}[c]{@{}c@{}}Format \\ Strategy\end{tabular} & Existing Works       & \begin{tabular}[c]{@{}c@{}}Runtime \\ Overhead\end{tabular} \\ \midrule
    Full-Graph-Level                                                     & Static                                                     & GNNAdvisor~\cite{GNNAdvisor}, NeuGraph~\cite{NeuGraph} & Low                                                         \\
    Block-Level                                                          & Adaptive                                                   & PCGCN~\cite{PCGCN}                & High                                                        \\
    \textbf{Subgraph-Level}                                                      & \textbf{Adaptive}                                                   & \textbf{Ours}                 & \textbf{Low}                                                         \\ \bottomrule
    \end{tabular}
    }
\end{table}

Table \ref{tbl:kernel_mapping} compares our system against existing GNN acceleration works.
Previous works either employ full-graph-level execution granularity, not fully exploiting the performance optimization opportunities provided by the distribution of intra-graph densities, or utilize block-level execution strategies that incur substantial runtime overhead. 
In contrast, \proj{} utilizes subgraph-level granularity with adaptive kernel mapping, effectively leveraging the performance optimization opportunities presented by both intra- and inter-graph density distributions while minimizing runtime overhead as much as possible.

\subsection{Subgraph-level Customized Kernel}

In this subsection, we describe the design of subgraph-level kernels in \proj{}. 
The optimization of graph-related operations on GPUs requires two key considerations.
Firstly, it is essential to implement an appropriate mapping from computation to the CUDA software abstractions such as CTA (Cooperative Thread Array), thread, etc., to fully leverage the computing resources of the GPU~\cite{cuda_programming}. 
Secondly, the GPU's memory hierarchy, including global memory, shared memory, and registers, exhibits different access overhead from large to small~\cite{gpu_memory}. To achieve lower access overhead and better performance, efficient memory management is crucial. 
As an architecture designed to process regular, continuous data, GPU is inefficient when processing irregular sparse data such as graphs~\cite{gnn_compute_survey, Gnnmark}. So, customized optimizations are required due to the irregular and sparse nature of GNNs.

As previously mentioned in \Sec{sec:motivation}, there is a diversity in preferred input formats for graphs with different density levels. Furthermore, intra- and inter-community subgraphs exhibit different density distribution properties. 
To harness these distinct features, we design a series of density-specific subgraph-level kernels.

\begin{figure}[t]
    \centering
    \includegraphics[width=0.98\linewidth]{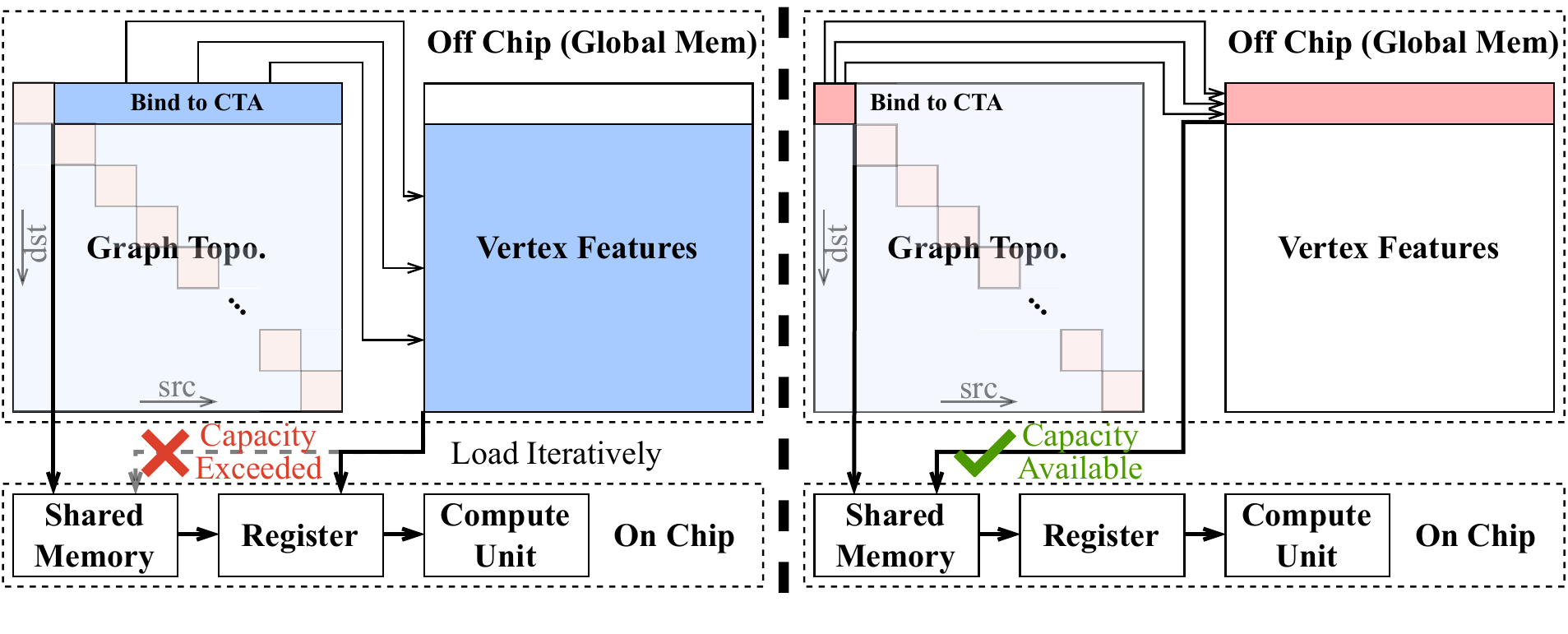}
    \caption{Comparison of kernel execution with CSR format for (left) low-density inter-community subgraph and (right) high-density intra-community subgraph.}
    \label{fig:csr_kernel_mapping}
    \vspace{-0.2cm}
\end{figure}

\paragraph{CSR-based kernel.}
As shown in the left side of \Fig{fig:csr_kernel_mapping}, we present the CSR-based kernel for inter-community subgraphs, which are characterized by low and irregular density distributions. 
Since the CSR format stores the adjacency matrix in a row-major way, our approach maps a CTA to multiple destination vertices, with each thread accessing the corresponding source neighbor's vertex features serially. To exploit the reuse of topological data within the CTA, we cache the data in shared memory. On the other hand, as the index of source neighbors cannot be determined within a finite range that fits the size of shared memory, we load the vertex features directly from global memory into registers.

Next, we introduce our CSR-based customized kernel for intra-community subgraphs. These subgraphs are characterized by high density, with edges concentrated in blocks around the diagonal of the adjacency matrix.
In light of this observation, we present our customized CSR-based kernel for intra-community subgraphs as shown in the right side of \Fig{fig:csr_kernel_mapping}.
We map a CTA to a community on a corresponding adjacency matrix.
In this mapping relationship between CTA and graph topology, the range of indexes of source vertices that a CTA needs to access is limited and can be determined in advance. Thus, we can pre-load all the vertex features that are required to be accessed into the shared memory.
This improves the efficiency of memory accesses, as the intra-community subgraph has a higher density and these features are accessed repeatedly.
Furthermore, to avoid excessive shared memory allocation from negatively affecting the parallelism of the CUDA kernel, we apply tiling techniques~\cite{cutlass} when the feature size is large.

\begin{algorithm}[b]
    \SetKwFor{ParallelFor}{for}{in parallel do}{end for}
    \SetKwData{Row}{Row}\SetKwData{Col}{Col}\SetKwData{ITensor}{X}\SetKwData{OTensor}{Y}
    \SetKwFunction{AtomicAdd}{Atomic\_Add}
    \SetKwInOut{Input}{input}\SetKwInOut{Output}{output}

    \Input{Graph $G=(Row[E], Col[E])$, Vertex Feature Tensor $X[V][F]$}
    \Output{Vertex Feature Tensor $Y[V][F]$}

    \ParallelFor{$edge\_id=0$ \KwTo $E-1$}{
        \ParallelFor{$dim\_id=0$ \KwTo $F-1$}{
            $dst\_id=$ \Row$[edge\_id]$\;
            $src\_id=$ \Col$[edge\_id]$\;
            \AtomicAdd{\OTensor$[dst\_id][dim\_id]$, \ITensor$[src\_id][dim\_id]$}\;
        }
    }
    \caption{The coo-based \texttt{aggregate-sum} kernel.}
    \label{algo:coo}
\end{algorithm}

\paragraph{COO-based kernel.}
The COO format is distinctive from the CSR format in that it organizes data in an edge-wise manner.
Therefore, we design COO-based kernel templates. As demonstrated by the example of the \texttt{aggregate-sum} kernel in \Algo{algo:coo}, this kernel performs computation by allocating threads and conducting element-wise computation. 
The topology and vertex feature data accessed in the COO kernel are independent between threads, and thus the shared memory caching mechanism is not employed.
This approach offers a high degree of parallelism, but destination vertices' updates must be atomic, making it more appropriate for input graph datasets with extremely low density. Consequently, the COO-based kernel is only utilized as a code candidate for inter-community subgraphs.

\paragraph{Dense-based kernel.}
We also propose a dense format-based approach for the intra-community subgraph.
Specifically, we map a CTA to a community's adjacency matrix block and then use sequential access and computation to perform sparse graph operations. 
For example, by storing the adjacency matrix in dense format and then directly performing in batched GEMM kernel~\cite{cublas_library} on it with the vertex feature, the result is equivalent to the \texttt{aggregation-sum} graph operator. 
Traditional graph computation methods do not utilize this approach due to the low density of graphs and the resulting large number of invalid accesses and computations. However, in the case of intra-community subgraphs with high density, this method can provide optimal performance gains in some scenarios.
The use of Tensor Core for 32-bit computation has been supported only since the beginning of the Ampere architecture. To ensure computational equivalence, we use Tensor Core in the A100 GPU and CUDA Core in the V100 GPU as compute units for the dense format kernel in our subsequent implementations. The mixed precision computational approach enables the utilization of Tensor Core on pre-Ampere architectures, and we leave this as future work.

\subsection{Adaptive Selector}
After customizing the kernels for high-performance computational optimization of input subgraphs with varying densities on the back-end, the preprocessing stage involves utilizing a community-based reordering tool to decompose the input graph into inter-community and intra-community subgraphs. 
Specifically, we iterate through each edge of the graph after reordering and calculate the block index of the adjacency matrix where the edge is located using the indexes of the source and destination vertices of the edge during the preprocessing stage. 
When the block index corresponding to the source vertex is equal to the block index corresponding to the destination vertex, it means that the edge is on the diagonal of the adjacency matrix, i.e., it belongs to the intra- community subgraph. Otherwise, this edge is added to the inter-community subgraph.

The final pivotal aspect in attaining efficient training is the core component: the adaptive selector. This selector employs a feedback-driven approach to determine the most suitable kernels for different input graphs.
Due to the limited number of subgraph-level customized kernels provided by \proj{}, specifically two for intra-subgraph and two for inter-subgraph, and the fact that GNN training requires hundreds or even thousands of iterations with a static topology graph~\cite{gnn_survey}, we adopt a feedback-based selection strategy.
In the first few iterations of GPU training, we use a monitor to collect the running time of each subgraph kernel, which is then fed back to the runtime scheduler as the basis for kernel selection in the following iteration.
Although this feedback collection process may cause some time loss due to monitoring, the performance losses incurred in the early iterations are considered insignificant in the overall context of the training process as evaluated in \Sec{subsec:eval_overhead}.

%% file: tex/implementation.tex
\DeclareFixedFont{\ttb}{T1}{cmtt}{bx}{n}{7} 

\definecolor{deepblue}{rgb}{0,0,0.5}
\definecolor{deepred}{rgb}{0.6,0,0}
\definecolor{deepgreen}{rgb}{0,0.5,0}

\newcommand\pythonstyle{\lstset{
    language=Python,
    morekeywords={self},
    keywordstyle=\ttb\color{deepgreen},
    emph={AdaptGear,AG},
    emphstyle=\ttb\color{violet}
}}

\lstnewenvironment{python}[1][]
{
\pythonstyle
\lstset{#1}
}
{}

\begin{figure}[t]
    \centering
    \begin{minipage}[t]{0.98\linewidth}
    \begin{python}
import AdaptGear as AG
import torch

# Create a GCN class.
class GCN(torch.nn.Module):
    def __init__(self,...):
        self.gcn = AG.GCNConv(in_feats, h_feats)
        ...
    def forward(self, x, inter_subg, intra_subg):
        x = self.gcn(x, inter_subg, intra_subg)
        ...
        
# Define a GCN model.
model = GCN(...)

# Loading graph dataset.
graph = AG.load_graph(graph_file)

# Reorder and decompose graph.
inter_subg, intra_subg = AG.graph_decompose(graph, method='METIS', comm_size=16)

# Run model.
predict_y = model(x, inter_subg, intra_subg)

# Compute loss and accuracy.
# Gradient backpropagation for training.
    \end{python}
    \end{minipage}

    \caption{Example of GCN model using \proj{}'s interfaces.}
    \label{fig:gcn_frontend}
\end{figure}

\section{Implementation}
\label{sec:impl}

In this section, we elaborate on the implementation details of \proj{}.
We start with the front-end programming interface and describe how the back-end integrates with our system.

\subsection{User-level Interface}
\proj{} utilizes PyTorch as the front-end to enhance programmability and user-friendliness. As an illustration, a representative GCN model is provided as an example for using \proj{} in \Fig{fig:gcn_frontend}.

\proj{} offers two types of interfaces. 
The first is for the computation of GNN, such as the \texttt{AG.GCNConv} in Line 7 of \Fig{fig:gcn_frontend}. 
The second interface is for the preprocessing of graphs, such as the \texttt{AG.graph\_decompose} in Line 19.
Upon examination, it is clear that \proj{} does not substantially deviate from the traditional GNN framework at the front-end level, but only adds a customized computational interface and an additional preprocessing step for graph decomposition. Consequently, it preserves good scalability and lowers the users' learning curve.

Another feature in \proj{} is the adaptive selector, which operates in a transparent manner to the user during training. This design choice eliminates the need for the user to manually choose the optimal selection strategy, thereby improving ease of use.

\begin{figure*}[t]
    \centering
    \subfloat[GCN on V100.]{
        \begin{minipage}[b]{0.50\linewidth}
            \centering
            \includegraphics[trim=0 0 0 0, clip, width=0.98\linewidth]{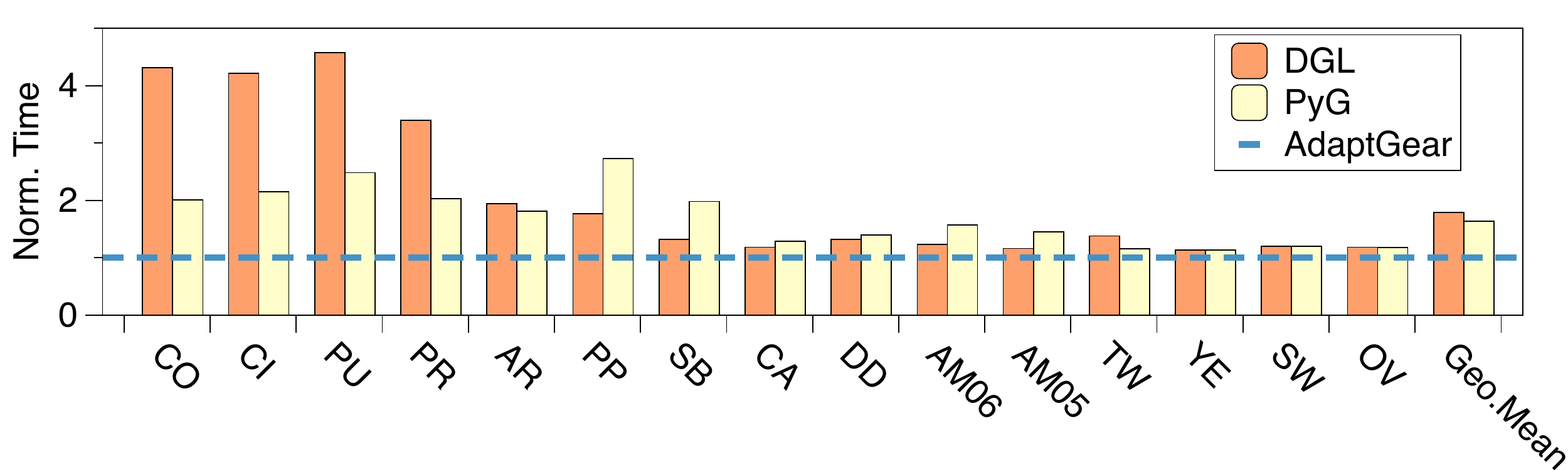}
        \end{minipage}
    }
    \subfloat[GIN on V100.]{
        \begin{minipage}[b]{0.50\linewidth}
            \centering
            \includegraphics[trim=0 0 0 0, clip, width=0.98\linewidth]{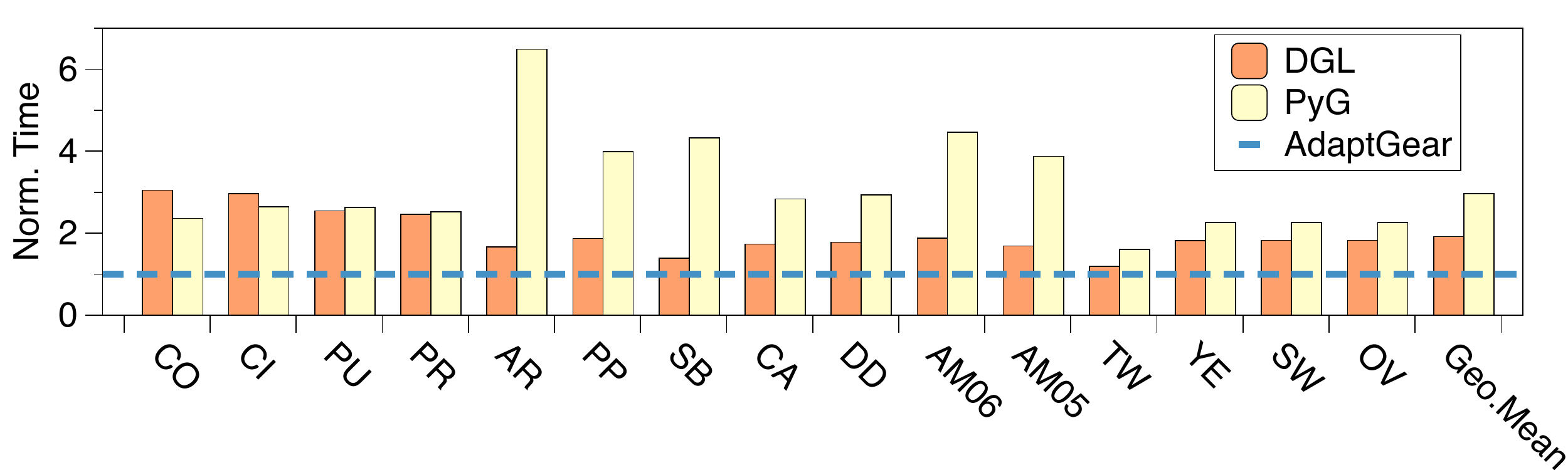}
        \end{minipage}
    }
    \vspace{-0.10cm}
    \medskip

    \subfloat[GCN on A100.]{
        \begin{minipage}[b]{0.50\linewidth}
            \centering
            \includegraphics[trim=0 0 0 0, clip, width=0.98\linewidth]{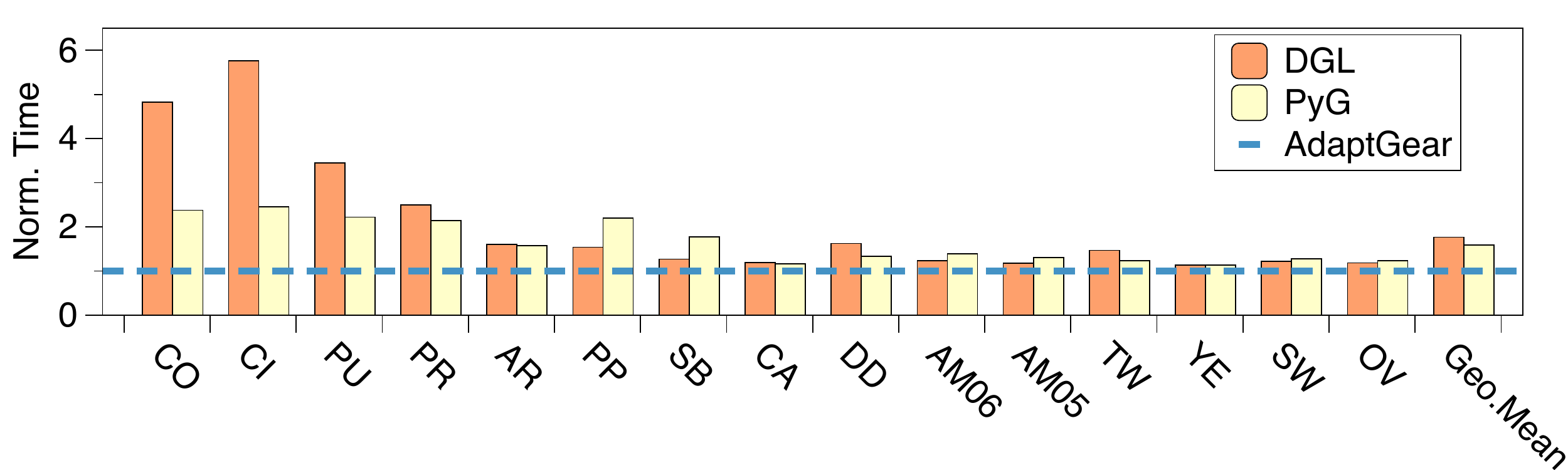}
        \end{minipage}
    }
    \subfloat[GIN on A100.]{
        \begin{minipage}[b]{0.50\linewidth}
            \centering
            \includegraphics[trim=0 0 0 0, clip, width=0.98\linewidth]{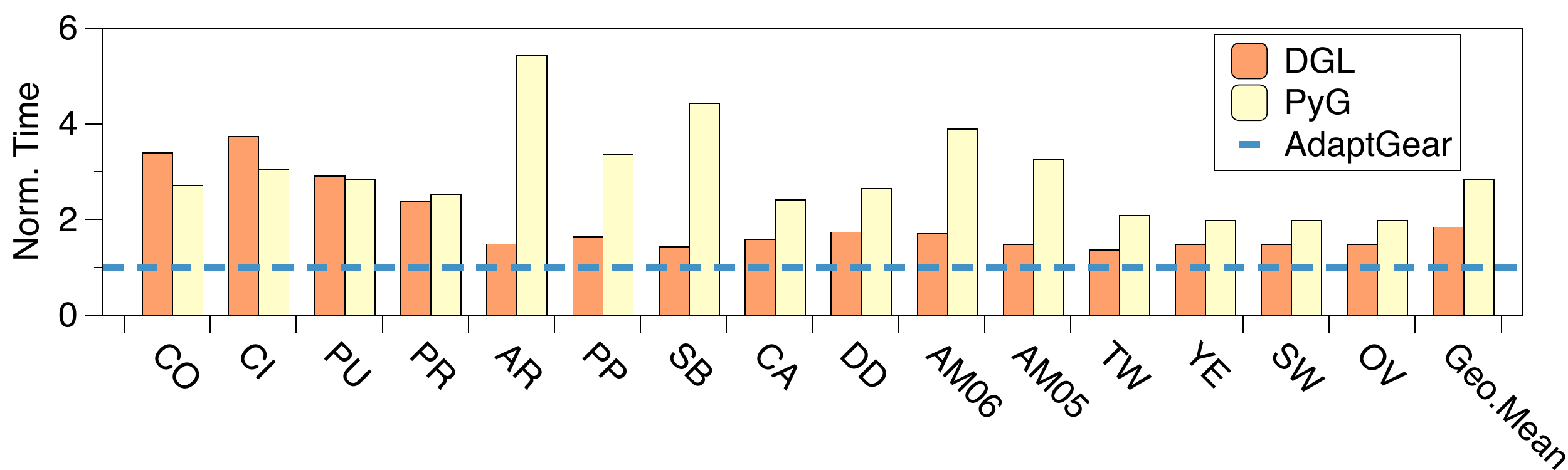}
        \end{minipage}
    }
    \caption{End-to-end normalized training time result on two GPUs. $X$-axis indicates different graph datasets.}
    \label{fig:e2e_dgl_pyg}
\end{figure*}
\subsection{System Integration}
In this subsection, we detail the integration of the front-end interface with the back-end implementation in \proj{}.

The front-end computational interface needs to be integrated with the back-end kernel implementation. \proj{} offers a variety of density-customized kernels, which are built using C++/CUDA and integrated into the PyTorch framework through pybind11~\cite{pybind11}.
The data is loaded using a Pytorch-based data loader and passed as a Tensor to \proj{}'s back-end for computation on GPUs. Upon completion of the computation, the result tensor is returned to the original PyTorch framework for further uses like loss computations.

The preprocessing interface in \proj{} consists of two stages: reordering and decomposition.
The first stage involves the graph reordering using the default algorithm, METIS~\cite{METIS}. This stage allows for the specification of either the community size or the number of communities as parameters, providing flexibility in the reordering process. 
Furthermore, the specific reordering algorithm used in the backend has potential for future expansion.
In the second stage, the graph is decomposed into intra- and inter-community subgraphs by traversing the graph once and dividing the edges based on their source and destination vertex indices.

%% file: tex/methodology.tex
\section{Methodology}
\label{sec:methodology}

\paragraph{Experiments} To evaluate \proj{}, we use two enterprise-level GPUs as our hardware platforms: Tesla V100~\cite{v100_whitepaper} and Ampere A100~\cite{a100_whitepaper}.
\Tbl{tbl:methodology} details our experimental setup.

\begin{table}[b]
    \caption{Detailed experimental setup.}
    \label{tbl:methodology}
    \resizebox{0.49\textwidth}{!}{
        \begin{tabular}{l|ll}
            \hline
            GPU      & \multicolumn{1}{l|}{NVIDIA Tesla V100(80 SMs)}                                                             & Nvidia Ampere A100 (108 SMs)                                                           \\ \hline
            CPU      & \multicolumn{1}{l|}{\begin{tabular}[c]{@{}l@{}}Intel(R) Xeon(R) Silver \\ 4210 CPU @ 2.20GHz\end{tabular}} & \begin{tabular}[c]{@{}l@{}}Intel(R) Xeon(R) Silver \\ 4210R CPU @ 2.40GHz\end{tabular} \\ \hline
            OS       & \multicolumn{1}{l|}{Ubuntu 18.04.5 (kernel 5.4.0)}                                                         & Ubuntu 20.04.2 (kernel 5.11.0)                                                         \\ \hline
            Driver       & \multicolumn{1}{l|}{GPU Driver: 470.57}                                                         & GPU Driver: 515.48.07                                                         \\ \hline
            Software & \multicolumn{2}{c}{\begin{tabular}[c]{@{}c@{}}CUDA: 11.6; Pytorch: 1.13\end{tabular}}                                                                  \\ \hline
            \end{tabular}
    }
\end{table}

\paragraph{Baselines}
We choose four baseline implementations for comparison: 1) Deep Graph Library (DGL)~\cite{DGL} is the state-of-the-art GNN framework that works for multiple DL frameworks. We choose PyTorch version in this work. 2) Pytorch-Geometric (PyG)~\cite{PyG} is another GNN framework which is built upon PyTorch. 3) GNNAdvisor~\cite{GNNAdvisor} accelerates GNNs on GPUs with handwritten full-graph-level CUDA kernel implementations. 4) PCGCN~\cite{PCGCN} utilizes block-level adaptive kernel to leverage the intra-graph hybrid density distribution to accelerate GCN.

\paragraph{Benchmarks}
We use GCN~\cite{GCN} and GIN~\cite{GIN} as representative GNN models in our study. We follow the default configuration for layers, hidden features, and training parameters as outlined in their original papers for all baselines and \proj{} to ensure a fair comparison. We run each benchmark 200 iterations of end-to-end training and present the average results to isolate the effects of randomness.

\paragraph{Datasets}
We use 15 graph datasets that have also been used in many previous GNN optimization works~\cite{sen2008collective,KKMMN2016,snapnets}.
The total count, sparsity, input feature size, and output classes vary significantly among these datasets. 
As such, our chosen datasets are sufficient to represent the graph in real-world scenarios.
\Tbl{tbl:dataset} provides detailed information for these datasets.

%% file: tex/eval.tex
\section{Evaluation}
\label{sec:evaluation}

In this section, we aim to evaluate the following points:
\begin{itemize}[ itemsep=0pt, parsep=0pt, labelsep=5pt, 
        leftmargin=*, topsep=0pt,partopsep=0pt]
\item What are the end-to-end performance improvements brought by \proj{} compared to existing GNN frameworks and manual optimization efforts?
\item How much does each design module
 in \proj{} contribute to the overall performance improvement?
\item How much is the additional overhead introduced by the design of \proj{}?
\end{itemize}

\subsection{End-to-end Performance Comparisons}

\paragraph{State-of-the-Art GNN Frameworks.}
We first compare the end-to-end training performance of \proj{} with two widely adopted GNN frameworks, including DGL~\cite{DGL} and PyG~\cite{PyG}.
The results, as shown in \Fig{fig:e2e_dgl_pyg}, compare their normalized end-to-end execution time on two GPUs. 
To ensure the fairness of comparison, we use the same METIS community size setting of 16 for both the baselines and \proj{}. Our results indicate that on both GPUs, \proj{} achieves significant performance improvements over the baselines, with geometric average speedup values of $1.83 \times$ and $2.16 \times$ over DGL and PyG, respectively. This improvement is due to \proj{}'s efficient exploitation of the density distribution at both the intra- and inter-graph levels. 
For each GNN model, \proj{} achieves an average improvement of $1.69 \times$ and $2.33 \times$ on GCN and GIN, respectively.
The more significant improvement on GIN is attributed to its higher proportion of time spent on graph-related operations, which are the main optimization scope of \proj{}.

\begin{figure}[t]
    \vspace{-0.3cm}
    \centering

    \subfloat[GCN.]
    {
            \includegraphics[trim=0 0 0 0, clip, width=0.98\linewidth]{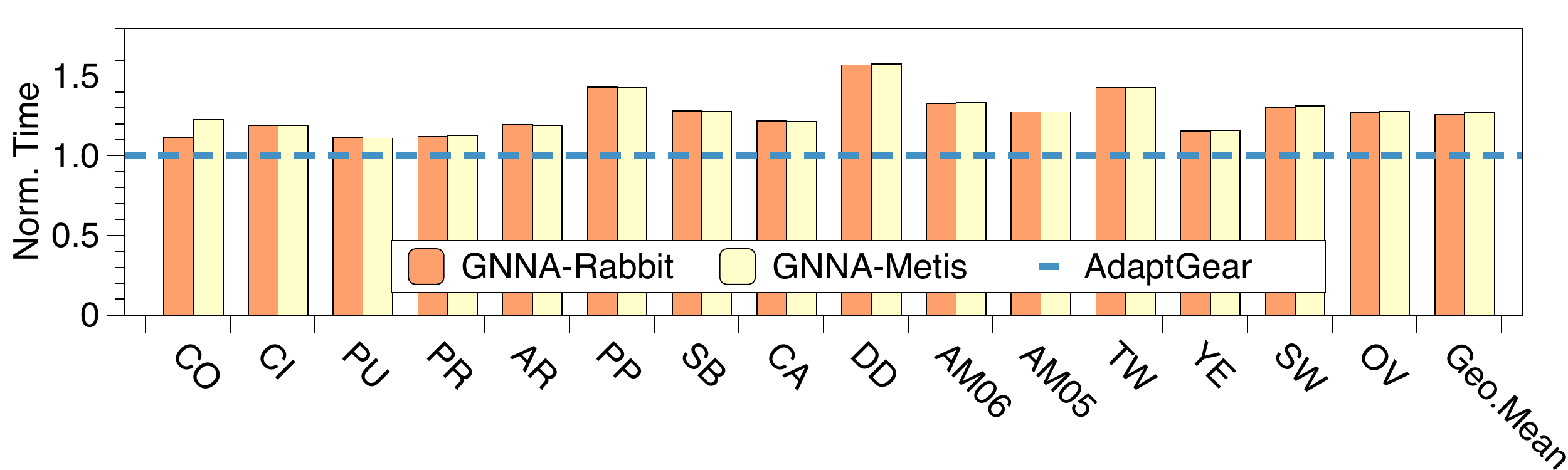}
        \label{fig:E2E_GNNA_PCGCN_A100_GCN}
    }
    \\
    \subfloat[GIN.]
    {
            \includegraphics[trim=0 0 0 0, clip, width=0.98\linewidth]{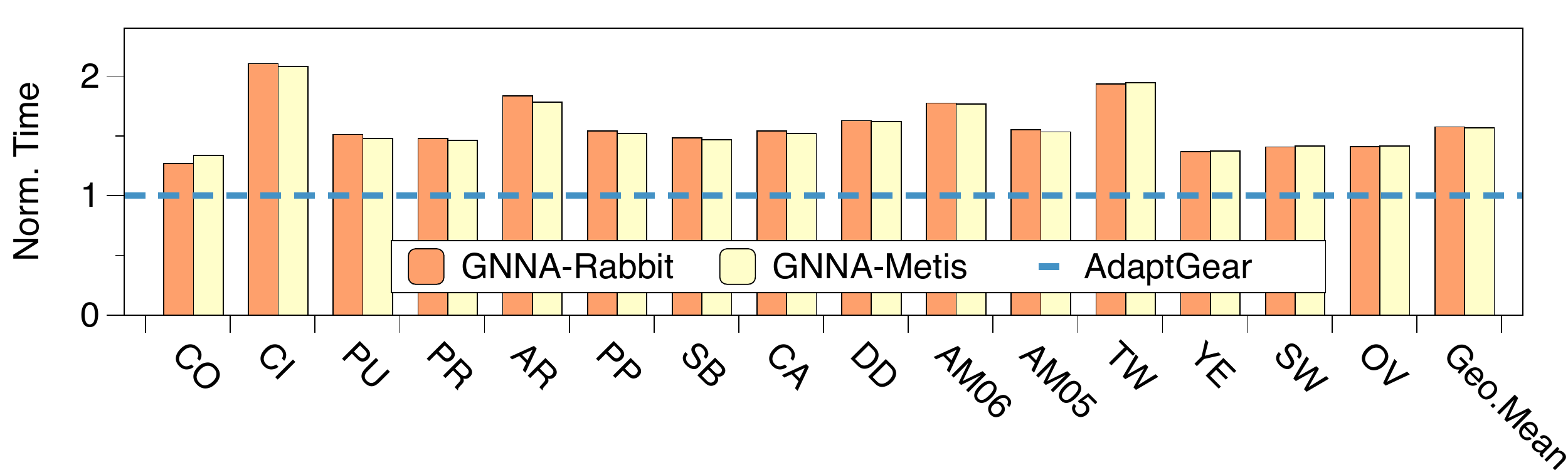}
        \label{fig:E2E_GNNA_PCGCN_A100_GIN}
   	}
    \caption{Performance comparisons between GNNAdvisor and \proj{} on A100 GPU. $X$-axis indicates different graph datasets.}
    \label{fig:eval_gnna}
\end{figure}

\paragraph{Manual Optimization.}
To demonstrate the performance benefits of \proj{} compared to existing GNN manual optimization efforts, we choose GNNAdvisor~\cite{GNNAdvisor} and PCGCN~\cite{PCGCN} as the full-graph-level and block-level acceleration baselines, respectively.

GNNAdvisor~\cite{GNNAdvisor} uses rabbit-sort~\cite{rabbit} as its default preprocessing step for reordering.
To eliminate the impact of the preprocessing tool, we collect performance results from two reordering preprocesses for GNNAdvisor, one using rabbit-sort (referred to as \texttt{GNNA-Rabbit}) and the other using METIS (referred to as \texttt{GNNA-Metis}).
Due to the space limit, we only show the results on the A100 GPU. 
As shown in \Fig{fig:eval_gnna}, \proj{} achieves an average $1.40 \times$ and $1.41 \times$ performance gain over \texttt{GNNA-Rabbit} and \texttt{GNNA-Metis} on A100 GPU.
This result shows that \proj{} yields good performance optimization results for different preprocessing methods.
It is worth noting that the results trend for the V100 GPU is similar. Specifically, the geometric average performance speedup ratios of \proj{} compared to \texttt{GNNA-Rabbit} and \texttt{GNNA-Metis} on the V100 GPU are $1.39 \times$ and $1.38 \times$, respectively.

\begin{figure}[t]
    \centering

    \includegraphics[width=0.98\linewidth]{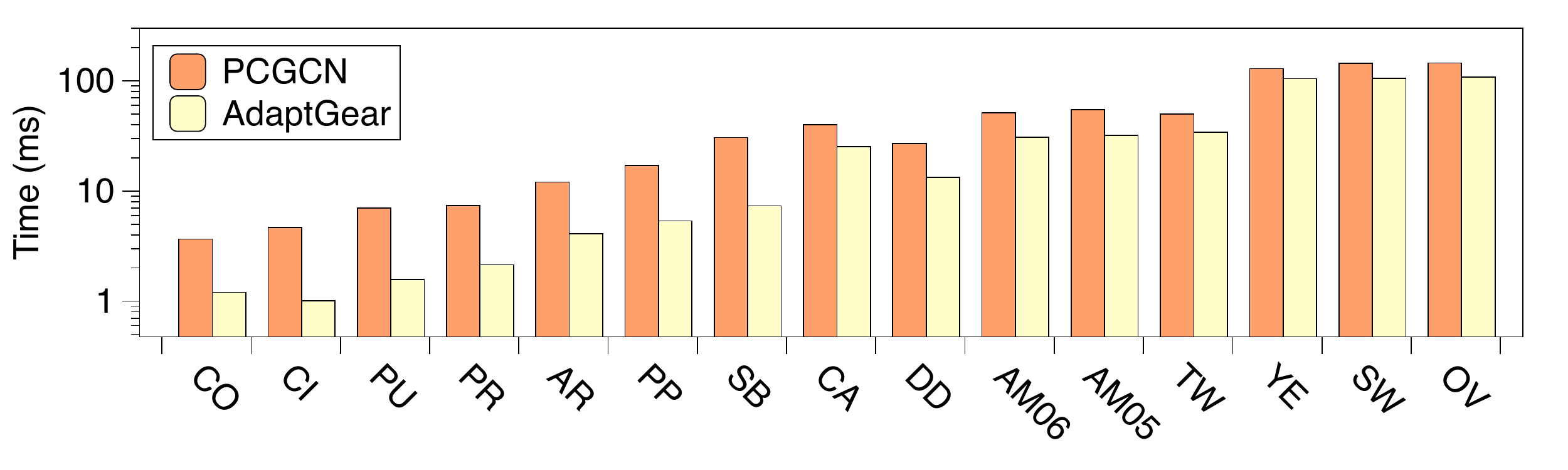}

    \caption{Performance comparisons of PCGCN and \proj{} for GCN on A100 GPU. $X$-axis indicates different graph datasets.}
    \label{fig:eval_pcgcn}
\end{figure}

PCGCN~\cite{PCGCN} optimizes GCN performance through a block-level approach. However, the METIS parameters used are not clearly specified in their paper. Therefore, we traverse the METIS parameters within the range of 2 to 1024 at multiples of 2 intervals for PCGCN and present the optimal one as the final performance results.
The results on A100, as shown in \Fig{fig:eval_pcgcn}, indicate that despite providing PCGCN with a wide range of reordering parameters, its performance remains lower compared to that of \proj{} for all datasets. Specifically, the geometric average performance speedup ratios of \proj{} over PCGCN on the A100 and V100 GPUs are $2.30 \times$ and $2.59 \times$, respectively.

Our comparison with these baselines highlights that the subgraph-level kernel mapping granularity utilized by \proj{} is the more effective for optimizing the computation in GNNs.

\subsection{Performance Improvement Breakdown}

\proj{} incorporates two performance optimization modules, including subgraph-level customized kernels and an adaptive selector.
To evaluate the effectiveness of these modules, we design relevant baselines and conduct experiments to quantify their contribution.
We define three optimization versions of \proj{}. The \texttt{O1} version utilizes a static CSR kernel at the full-graph level, while the \texttt{O2} version employs static CSR kernels for intra-community subgraphs and COO kernels for inner-community subgraphs. The \texttt{O3} version incorporates subgraph-level adaptive sparse kernels.

\begin{figure}[t]
    \centering
    \includegraphics[width=0.98\linewidth]{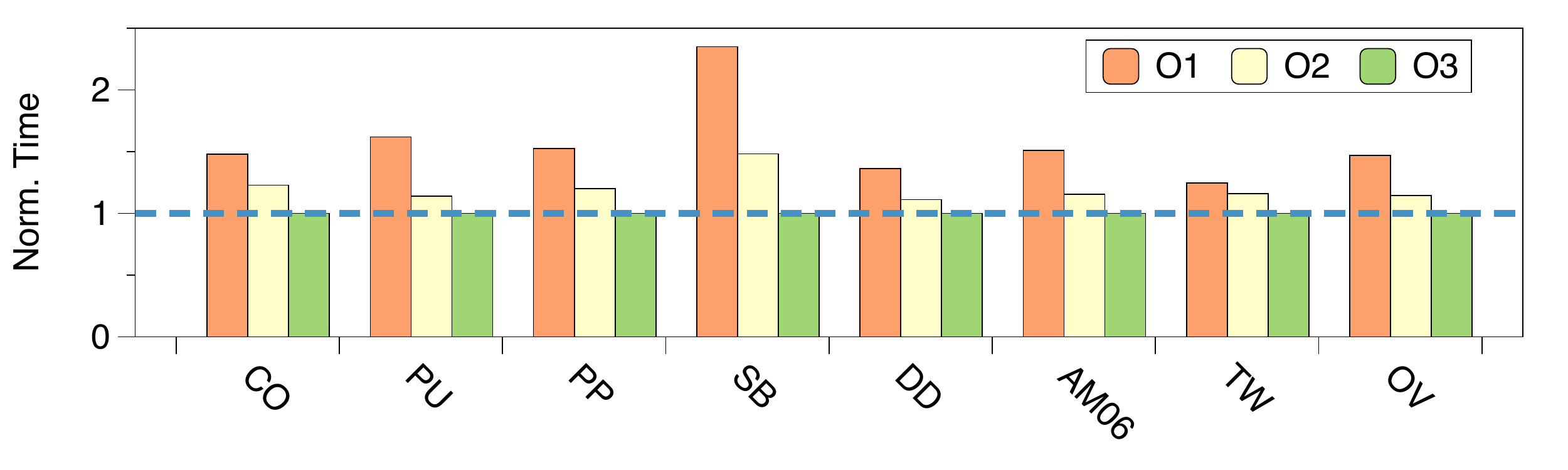}

    \vspace{-0.1cm}
    \caption{Execution time of different \proj{} versions on A100.}
    \label{fig:breakdown_opt}
    \vspace{-0.1cm}
\end{figure}

We use the GCN model in this experiment and collect the performance results on A100 GPU. Due to the space limit, we omit the results for the other models and hardware.  
As illustrated in \Fig{fig:breakdown_opt}, the results show that the implementation of different optimization versions can lead to significant variations in performance improvement across different datasets. This highlights the crucial role that subgraph-level kernels and the adaptive selector play in the optimization process of \proj{}.

\subsection{Additional Studies}
\label{subsec:eval_overhead}

\paragraph{Runtime Overhead.}
There are two sources of additional runtime overhead in \proj{}.
The first is graph preprocessing, which includes graph reordering and graph decomposition, and needs to be performed only once before the training starts. The second is the adaptive selector, which monitors performance in the early iterations of GNN training.
However, the overhead from both processes has a negligible impact compared to the overall GNN training process.
To illustrate, we evaluate the overhead using the amazon0601 dataset. The overhead of the graph decomposition is 0.08 seconds, while the overhead of the graph reordering is 0.59 seconds. 
Furthermore, the runtime monitoring in our adaptive selector incurs less than 0.1 seconds in the early training iterations.
Overall, the magnitude of these overheads is insignificant compared to the hours or even days required~\cite{gnn_survey, gnn_benchmarking} for the GNN training.

\begin{figure}[t]
    \centering
    \includegraphics[width=0.98\linewidth]{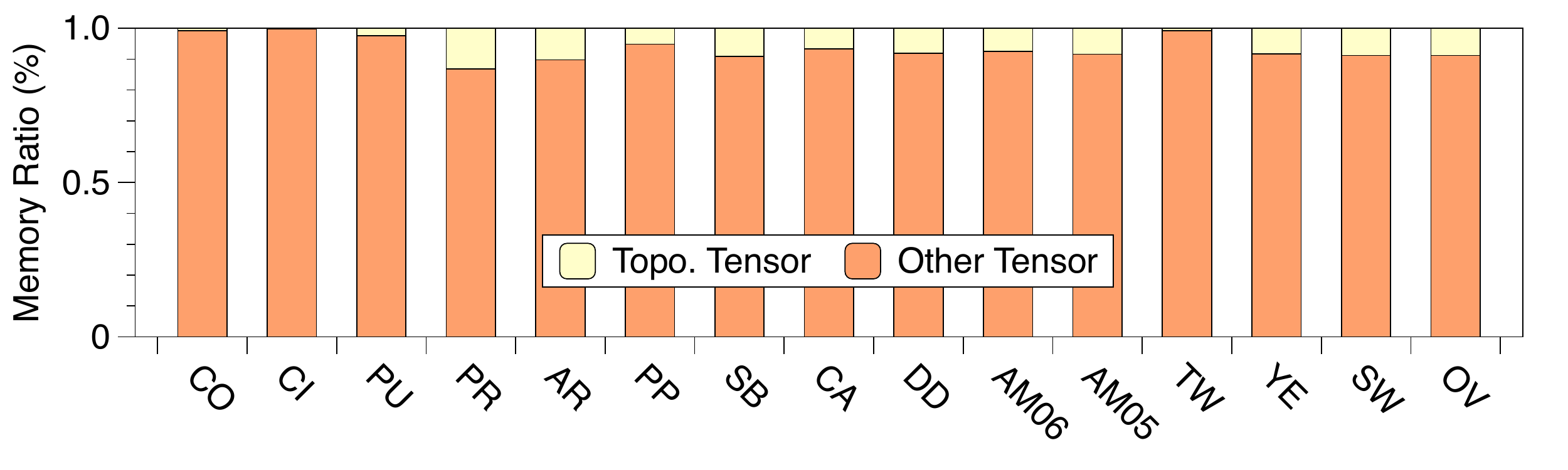}

    \caption{Memory overhead for storing subgraph topology (denoted as Topo. Tensor) in different datasets.}
    \label{fig:memory_overhead}
\end{figure}

\paragraph{Memory Overhead.}
The graph decomposition potentially generates additional storage overhead for the topology.
However, this overhead is comparatively small as the majority of memory usage during GNN training is occupied by vertex features and gradient-related data~\cite{zhou2021brief}.
To quantify the additional memory overhead, we measure the maximum memory overhead via PyTorch Profiler~\cite{pytorch} for the GCN model on A100 GPU and compare it with the overhead for the additional subgraph topology storage. As shown in \Fig{fig:memory_overhead}, the percentage of memory occupied by additional topological data is only 4.47\% on average, which we believe is acceptable.

%% file: tex/related.tex
\section{Related Work} 

\paragraph{Graph Processing on GPUs.}
GPUs are crucial computing platforms in various task scenarios~\cite{kong2011mechanism,ma2017transt}.
Many studies have focused on optimizing GPUs at the architecture~\cite{guo2020balancing,wang2021dual,guo2022ant,leng2020asymmetric}, scheduling~\cite{guo2022nesting,wang2017quality,zhou2017reinforcement,liu2022veltair,cui2022dvabatch,cui2021enable}, and algorithm~\cite{guo2020accelerating,zhou2021characterizing,guo2022squant} levels.
Numerous graph processing systems~\cite{Cusha, PowerGraph, Tigr, Gunrock, graphchi, Mizan, Xstream} have been proposed to accelerate traditional graph algorithms on GPUs. 
These efforts have also explored various parallelization strategies, including vertex parallelism and edge parallelism for graph processing.
There are also efforts to explore dynamic parallelization strategies through domain-specific language (DSL)~\cite{GraphIt, GraphItGPU}.

However, GNNs differ from traditional graph algorithms in terms of their graph operation characteristics and the dimensionality of their feature embeddings, leading to a parallelization strategy space that exceeds the capabilities of traditional graph processing systems.

\paragraph{GNN Frameworks.}
DGL~\cite{DGL} and PyG~\cite{PyG} are two popular GNN frameworks, which both employ a message-passing programming interface based on DNN frameworks.
$G^3$~\cite{G3} focuses on utilizing graph processing frameworks for training GNNs on GPUs.
Despite these efforts, the existing frameworks do not fully exploit the opportunities presented by graph density characteristics for optimizing graph kernel computations and hence cannot achieve optimal performance on GPUs.

On the other hand, other frameworks such as Roc~\cite{Roc}, NeuGraph~\cite{NeuGraph}, and AliGraph~\cite{Aligraph} aim to address large-scale distributed GNN processing. However, their designs are orthogonal to \proj{}.
For handling large graphs requiring multiple GPUs to process, various well-established graph partitioning techniques can divide the graph into smaller subgraphs suitable for single-GPU training~\cite{METIS,balance_partition,multilevel_partition}. Therefore, the optimization of single-GPU training is equally beneficial for multi-GPU scenarios.

\paragraph{Graph Kernel Optimization.}
There are also some works that try to explore the graph kernels in GNNs to optimize GNNs. 
GNNAdvisor~\cite{GNNAdvisor} provides customized kernels with scalable parameters for accelerating GNNs on GPUs.
GE-SpMM~\cite{Ge-SPMM} focuses on optimizing SPMM-like graph kernels in GNNs, while FeatGraph~\cite{FeatGraph} extends TVM~\cite{tvm} to execute SPMM-like and SDDMM-like graph kernels on GPUs.
uGrapher~\cite{zhou2023ugrapher} addresses the challenges posed by the dynamics of input graphs and operators by providing a unified abstraction for all graph operations.
However, the previous optimization efforts apply to entire input graphs and overlook the performance optimization opportunities presented by variations in the density distribution of intra-graphs.
Additionally, PCGCN~\cite{PCGCN} adopts a block-level approach to leverage the intra-graph density distribution. However, it leads to additional accumulation operations, resulting in high runtime overhead in some scenarios.

Unlike all existing work, \proj{} propose a set of subgraph-level customized kernels and adaptively select the current most suitable kernel for a given input, thus achieving a general GNN high-performance computation optimization.

%% file: tex/conclusion.tex
\section{Conclusion}

In this work, we propose \proj{}, a novel high-performance GNN training system that exploits both the intra- and inter-graph sparsities via adaptive subgraph-level kernels.
Our system achieves a significant average speedup of $1.87 \times$ compared to various state-of-the-art works, including GNN frameworks and manual optimizations. 
The exceptional performance of \proj{} is attributed to two key components: subgraph-level customized kernels that offer diverse kernel formats optimized for subgraphs of varying densities and an adaptive selector that selects the appropriate kernel for the input subgraph.
Through the detailed evaluation, we demonstrate the effectiveness of \proj{} and its potential as a common methodology for high-performance GNN training.

\begin{acks}
This work was supported by the National Key R\&D Program of China under Grant 2021ZD0110104, the National Natural Science Foundation of China (NSFC) grant (62222210, U21B2017, and 620722-97).
The authors would like to thank the anonymous reviewers for their constructive feedback for improving the work. 
Any opinions, findings, and conclusions in this paper are those of the authors only and do not necessarily reflect the views of our sponsors.
\end{acks}